\documentclass{ws-procs9x6-cpt16}
\begin{document}

\newcommand{\refeq}[1]{(\ref{#1})}
\def\etal {{\it et al.}}

\title{Development of a low-energy, high-brightness $\mu^+$ beam line}

\author{A.\ Eggenberger,$^{1,\ddag}$ I. Belosevic,$^1$ and G. Wichmann$^1$}

\address{$^1$Institute for Particle Physics, ETH Zurich,\\
8038 Zurich, Switzerland\\
$^{\ddag}$E-mail: egandrea@phys.ethz.ch}

\author{On behalf of the muCool Collaboration\footnote{ \url{http://www.edm.ethz.ch/research/muoncooling.html}}}

\begin{abstract}
We are developing a beam line which compresses the phase space of a standard surface $\mu^+$ beam by 10 orders of magnitude with an efficiency of $10^{-3}$. Phase space compression occurs in a He gas target and consists of three consecutive stages: Transverse (perpendicular to the beam axis) compression, longitudinal compression and re-extraction into vacuum. Transverse compression was observed for the first time and longitudinal compression has been measured to occur within 2.5~$\mu$s with high efficiency.
\end{abstract}

\bodymatter

\section{Introduction}

The muon presents a unique probe for a wide field of research such as the muon g-2 experiment, $\mu$SR measurements,  muonium (Mu=$\mu^+ e^-$) spectroscopy or searches for Mu-$\overline{\text{Mu}}$ conversion.\cite{Gorringe2015} Additionally, both the muon and muonium atom are suited to search for Lorentz and CPT violation.\cite{Kostelecky2014} Moreover, the muonium atom allows to measure the gravitational acceleration of antimatter.\cite{Kaplan2016} Common to all these experiments is the demand for a high-brightness, low-energy $\mu^+$ beam. 

Conventional surface $\mu^+$ beams have an energy of 4.1~MeV, are close to 100\% polarized but posses a large phase space. In order to efficiently improve the phase space quality, a fast cooling method is needed to avoid substantial losses due to the short $\mu^+$ lifetime of 2.2~$\mu$s. We are currently developing a phase space compression scheme that stops a $\mu^+$ beam in a few mbar of He gas. By means of strong electric and magnetic fields the stopped $\mu^+$ swarm is compressed sequentially in transverse (perpendicular to the incoming beam axis) and longitudinal direction (along the beam and the B-field) into a small volume.\cite{Taqqu2006,Bao2014} The small $\mu^+$ swarm is then extracted through a 1~mm orifice into vacuum and re-accelerated. The full compression occurs within less than 10~$\mu$s, yielding a phase space reduction of $10^{10}$ with larger than $10^{-3}$ efficiency, dominated by the muon decay. All three compression stages can be tested individually.

\section{Basic Idea}

The drift velocity vector $\vec{v}_D$ of charged particles (here $\mu^+$) in a gas can be written as\cite{Lohse1992}
\begin{equation}
\vec{v}_{D} = \frac{\mu E}{1 + \omega^{2}\tau_{c}^{2}} \left[ \hat{E} + \omega \tau_{c} \left( \hat{E} \times \hat{B} \right) + \omega^{2} \tau_{c}^{2} (\hat{E} \cdot \hat{B}) \hat{B} \right],
\label{EqDriftVelVec}
\end{equation}
where $\hat{E}$ and $\hat{B}$ are the unit vectors along $\vec{E}$ and $\vec{B}$, $\mu$ the muon mobility, $\omega$ the cyclotron frequency and $\tau_c$ the mean time between collisions. Depending on $\tau_c$, i.e. the gas density, different terms in Eq.\ \refeq{EqDriftVelVec} can be made dominant, which is the key requisite for our compression scheme. 

\begin{figure}[htbp]
	\begin{minipage}[hbt]{0.53\textwidth}
		\includegraphics[width=1\textwidth]{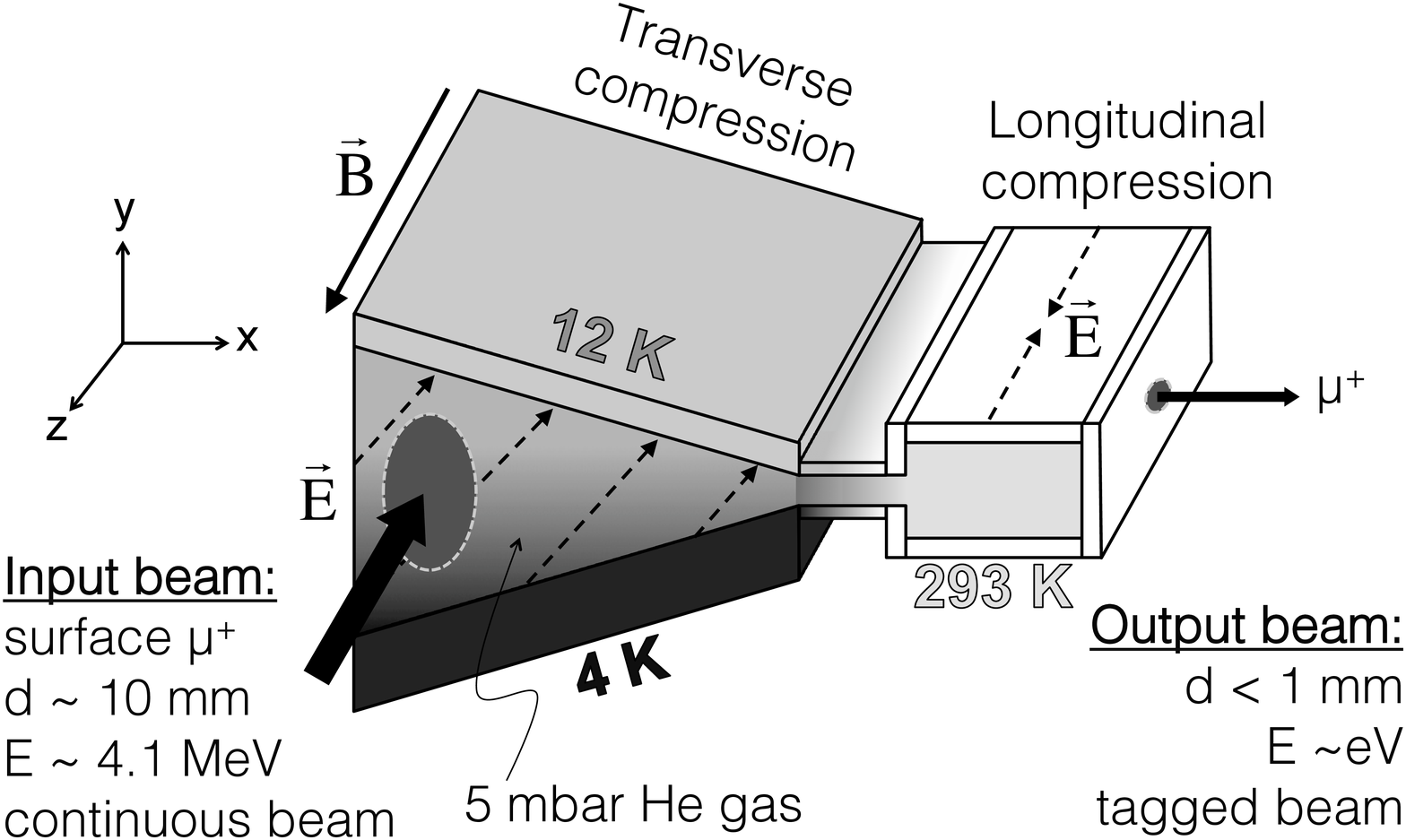}
	\end{minipage}
	\hfill
	\begin{minipage}[hbt]{0.43\textwidth}
		\includegraphics[width=1\textwidth]{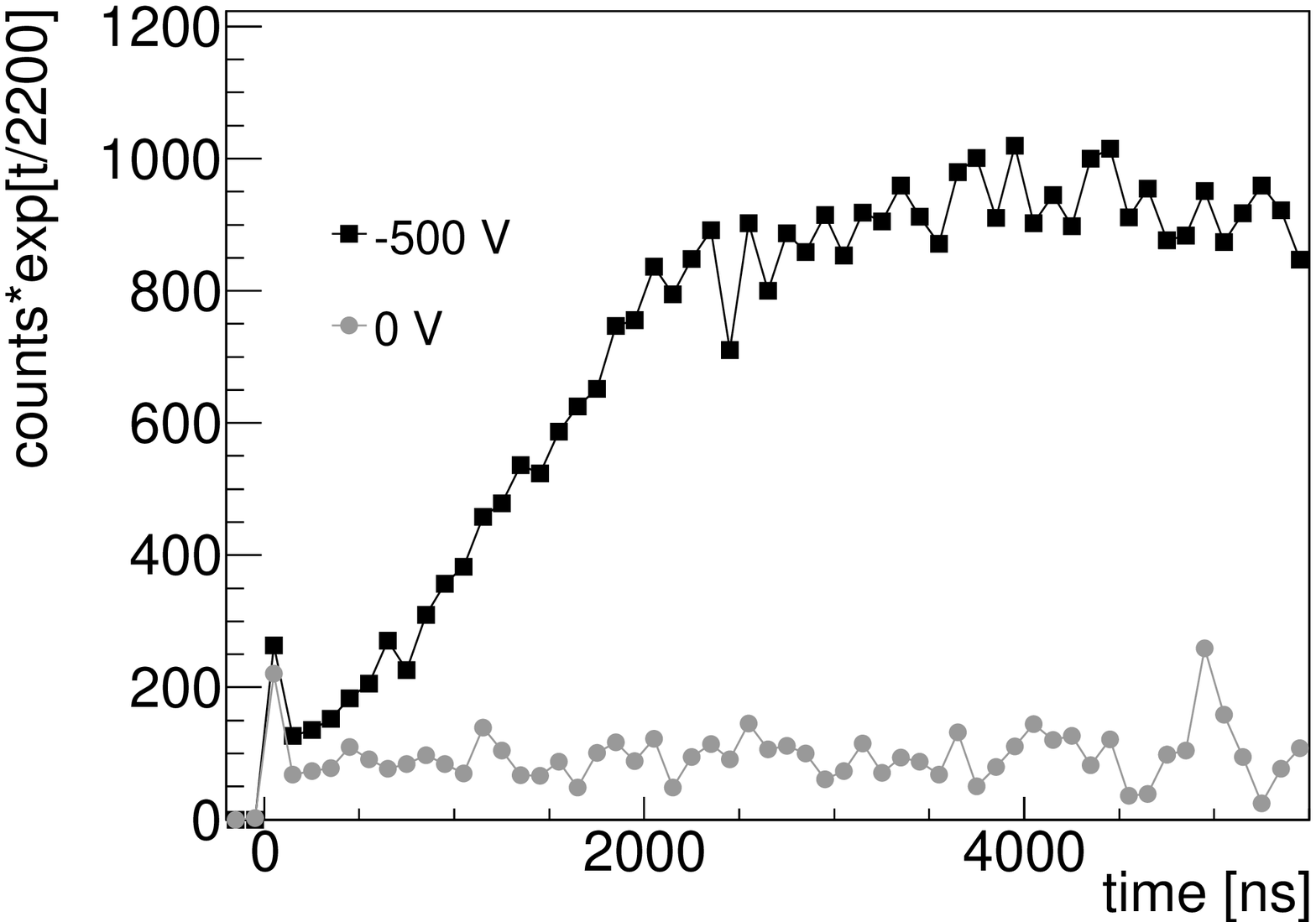}
	\end{minipage}
	\caption{(Left) Schematic for the phase space compression. $\mu^+$ are stopped in 5~mbar He gas at cryogenic temperatures inside a 5~T magnetic field. First, transverse compression occurs due to a gas density gradient and suitable electric fields, followed by longitudinal compression and extraction into vacuum. (Right) Time spectra of positron counts in the center of the longitudinal compression target. Applying an attractive negative potential (squares) compresses the $\mu^+$ towards the center, which is reflected in the increase of positron counts at later times compared to no applied potential (circles).}
	\label{Fig1}
\end{figure}

Figure\ \ref{Fig1} (left) illustrates the basic concept of our compression target. The incoming $\mu^+$ beam is stopped in a cryogenic gas target containing 5~mbar of He gas. Strong electric fields are applied perpendicular to the 5~T magnetic field: $\vec{B}=(0,0,5)$~T and $\vec{E}\approx(1,1,0)$~kV/cm. A stationary temperature gradient is created by cooling the lower part to 4~K while heating the upper part to 12~K.\cite{Wichmann2016} The resulting gas density gradient introduces a position-dependence of $\vec{v}_D$, thus $\mu^+$ stopped in the lower (upper) part of the target drift along the direction of $\hat{E}$ ($\hat{E} \times \hat{B}$), compressing the stop distribution in vertical ($y$)-direction.

The second compression stage is at room temperature. Because of the low He gas density, the last term in Eq.\ \refeq{EqDriftVelVec} dominates ($\omega \tau_c\gg1$). An electric field along the $z$-axis points towards the center of the target, leading to a compression in $z$-direction. A non-vanishing field component in $y$-direction steers the compressed muon swarm towards the point of extraction, where a final compression in both $y$- and $z$-directions occurs before extraction into vacuum through a small orifice. 

\section{Experimental Status}

The target for transverse compression meeting the various technical demands has been developed. In 2015 we could demonstrate the feasibility and obtain efficient compression. Data analysis is ongoing. The longitudinal compression was demonstrated in 2014. A 11~MeV/c $\mu^+$ beam entered the longitudinal target. In the center of the target a negative electric potential was applied, whereas the outer parts were kept at ground. The fraction of $\mu^+$ stopped in 5~mbar of He gas were subject to the resulting attractive force towards the target center.  Two coincidence detectors allow to construct a time spectrum from the detected decay positrons, as shown in Fig.\ \ref{Fig1} (right). If  a negative (attractive) potential is applied the number of coincident positrons increases, compared to the case without electric field (circles), demonstrating $\mu^+$ compression. A large compression efficiency of at least $>50$\% (subject to ongoing data analysis) is observed within less than 2.5~$\mu$s, not accounting the muon decay. 

\section{Conclusion}

We have demonstrated longitudinal and transverse compression in separate test experiments. Data analysis is ongoing, and we are starting to construct the transition region between the transverse and the longitudinal stages, and the differential pumping section for the extraction into vacuum. We acknowledge funding by the SNF \#200020$\_$159754.


\begin{thebibliography}{xx}
	\bibitem{Gorringe2015}
	T.P.\ Gorringe \etal,
	Prog.\ Part.\ Nucl.\ Phys.\ {\bf 84}, 73 (2015).
	\bibitem{Kostelecky2014}
	A.H.\ Gomes \etal,
	Phys.\ Rev.\ D\ {\bf 90}, 076009 (2014).
	\bibitem{Kaplan2016}
	D.M.\ Kaplan, these proceedings.
	\bibitem{Taqqu2006}
	D.\ Taqqu, Phys.\ Rev.\ Lett.\ {\bf 97}, 194801 (2006).
	\bibitem{Bao2014}
	Y.\ Bao \etal, Phys.\ Rev.\ Lett. \ {\bf 112}, 224801 (2014).
	\bibitem{Lohse1992}
	T.\ Lohse \etal, Adv.\ Ser.\ Direct.\ High Energy Phys.\ {\bf 9}, 81 (1992).
	\bibitem{Wichmann2016}
	G.\ Wichmann \etal, Nucl.\ Instr.\ Meth.\ Phys.\ Res.\ A {\bf 814}, 33 (2016).

\end{thebibliography}
\end{document}